\documentclass[fleqn,usenatbib]{mnras}
\usepackage{newtxtext,newtxmath}
\usepackage{comment}
\usepackage{multirow}
\usepackage{mathtools}
\usepackage[T1]{fontenc}
\usepackage{ae,aecompl}

\usepackage{graphicx}	
\usepackage{amsmath}	
\usepackage[ruled,vlined]{algorithm2e}
\usepackage{algpseudocode}
\usepackage{enumitem} 
\usepackage{xfrac} 

\setlist[itemize]{leftmargin=*}

\newcommand{\zsol}[1]{{#1}\:\mathrm{Z_\odot}}
\newcommand{\msol}[1]{{#1}\:\mathrm{M_\odot}}
\newcommand{\Sigmasol}[1]{{#1}\:\mathrm{M_\odot\:pc^{-2}}}
\newcommand{\rhosol}[1]{{#1}\:\mathrm{M_\odot\:pc^{-3}}}

\newcommand{\sevn}{\texttt{SEVN}}

\newcommand{\bifrost}{\texttt{BIFROST}}

\title[Hierarchical SMBH seed formation \& GWs]{{A rapid channel for the collisional formation and gravitational wave driven mergers of supermassive black hole seeds at high redshift}}
\author[A. Rantala et al.]{Antti Rantala$^{1}$\thanks{E-mail: anttiran@mpa-garching.mpg.de}, Thorsten Naab$^{1}$\\
$^{1}$Max-Planck-Institut f\"ur Astrophysik, Karl-Schwarzschild-Str. 1, 
D-85748, Garching, Germany\\
}
\date{Accepted XXX. Received YYY; in original form ZZZ}
\pubyear{2025}

\begin{document}
\label{firstpage}
\pagerange{\pageref{firstpage}--\pageref{lastpage}}
\maketitle

\begin{abstract}
Motivated by JWST observations of dense, clumpy and clustered high redshift star formation, we simulate the hierarchical assembly of nine $M_{\mathrm{cl}}=\msol{10^6}$ star clusters using the \bifrost{} N-body code. Our low metallicity models ($0.01Z_\odot$) with post-Newtonian equations of motion for black holes include evolving populations of single, binary and triple stars. Massive stars grow by stellar collisions and collapse into intermediate mass black holes (IMBHs) up to $M_\mathrm{\bullet}\sim\msol{6200}$, stellar multiplicity boosting the IMBH masses by a factor of $2$--$3$. The IMBHs tidally disrupt (TDE) $\sim50$ stars in $10$ Myr with peak TDE rates up to $\Gamma\sim5\times10^{-5}$ yr$^{-1}$ per cluster. These IMBHs are natural seeds for supermassive black holes (SMBHs) and the hierarchical assembly frequently leads to $>2$ SMBH seeds per cluster and their rapid mergers ($t<10$ Myr). We propose that a gravitational wave (GW) driven merger of IMBHs with $\msol{1000} \lesssim M_\bullet \lesssim \msol{10000}$ with comparable masses is a characteristic GW fingerprint of SMBH seed formation at redshifts $z>10$, and IMBH formation in similar environments at lower redshifts. Massive star clusters provide a unique environment for the early Universe GW studies for the next-generation GW observatories including the Einstein Telescope and the Laser Interferometer Space Antenna.
\end{abstract}



\section{Introduction}

Gravitational wave (GW) driven black hole (BH) mergers in the intermediate-mass black hole (IMBH; \citealt{Mezcua2017,Greene2020}) mass range of $M_\bullet \sim \msol{10^2}$--$\msol{10^5}$ at redshifts $z>6$ will be observable with the next-generation GW observatories such as the Laser Interferometer Space Antenna (LISA; \citealt{Amaro-Seoane2017}), the Einstein Telescope (\citealt{Punturo2010}) and the Cosmic Explorer (\citealt{Reitze2019}). This high-z GW window provides an unprecedented opportunity to constrain the theoretical models for the origins of supermassive black holes (SMBHs), a major open question in modern astrophysics only deepened by the James Webb Space Telescope (JWST) observations of accreting SMBHs in active galactic nuclei (AGN) at $z>10$ \citep{Maiolino2024}. To explain the $z=6$--$7$ AGN population \citep{Fan2023,Inayoshi2020}, $M_\bullet = \msol{10^4}$ SMBH seeds growing at the Eddington rate must have formed around $z\sim10$--$11$ to reach $M_\bullet = \msol{10^9}$ by $z=6$. By similar argument the progenitor of the GN-z11 SMBH ($M_\bullet \sim \msol{1.6\times10^6}$; \citealt{Maiolino2024}) should have formed at $z\sim18$--$19$ if it initially had $M_\bullet = \msol{10^4}$.

Out of the various proposed scenarios for the SMBH seed formation \citep{Rees1984}, the collisional runaway (e.g. \citealt{Gold1965, Spitzer1966}) in dense and massive star clusters is an attractive scenario for the SMBH origins as it relies on relatively well understood physical processes and requires little fine tuning. Recent observations by the JWST have revealed dense, clumpy star formation as well as bound $\gtrsim \msol{10^6}$ sub-pc size star clusters at $6\lesssim z \lesssim 10$ (e.g. \citealt{Fujimoto2024,Adamo2024}). Furthermore, observed high-z nitrogen enrichment at low metallicities and high star formation densities suggest the presence of dense environments in which stellar collisions can frequently occur (e.g. \citealt{Topping2024,Isobe2025}). This view of dense and clustered low metallicity star formation is further supported by Milky Way archaeology \citep{Belokurov2022}. Moreover, high-resolution or star-by-star hydrodynamical simulations of low metallicity clouds \citep[see e.g.][]{2018NatAs...2..725H,Polak2024b} and galactic star-bursts \citep{Lahen2020} indicate that massive star cluster formation is an inherently hierarchical process. Indeed, the regions of most intense star formation in the local Universe such as R136 in the Large Magellanic Cloud \citep{Fahrion2024} and in the Antennae galaxies \citep{Whitmore2010} host complex, hierarchical assemblies of star clusters, not isolated monolithically collapsed objects. Still, most numerical work on the stellar runaway scenario (e.g. \citealt{PortegiesZwart1999, PortegiesZwart2004, Gurkan2004, Freitag2006a, Mapelli2016, Sakurai2017, Rizzuto2021, ArcaSedda2023a, GonzalezPrieto2024, Vergara2025}) has focused on isolated, spherical star cluster models or idealised few-cluster setups \citep{Fujii2014}. Very recent models have introduced collisional stellar dynamics into star-by-star hydrodynamical simulations (e.g. \citealt{Fujii2024,Polak2024a,Cournoyer-Cloutier2024,Lahen2025}).

In \cite{Rantala2024b} we showed using direct N-body simulations that the hierarchical star cluster assembly with single stars boosts IMBH formation compared to monolithically formed clusters of the same mass and final structure. Massive stars can grow up to $\sim \msol{2000}$ through stellar collisions, collapse into IMBHs and further grow by tidal disruption events (TDEs) and GW driven BH mergers. In this Letter, present a new suite of nine hierarchical N-body simulations of massive star cluster formation including initial (primordial) populations of binary and triple stars. Based on our simulation results we propose that a close to equal mass GW merger in the IMBH mass range is a characteristic GW fingerprint of the hierarchical runaway collisional scenario for the SMBH seed formation. Crucially, rapid IMBH-IMBH mergers frequently occur within $<10$ Myr of the cluster formation.

This Letter is structured as follows. We present our N-body simulation code \bifrost{} and the numerical setup in section 2. In sections 3 and 4 we report our main results, focusing on stellar collisions and GW mergers. Finally, we summarise in section 5. The structure, kinematics and the stellar multiplicity content of the assembled star clusters is analysed in detail in a separate study \citep{Rantala2025b}.

\begin{figure}
\begin{centering}
\includegraphics[width=0.6\columnwidth]{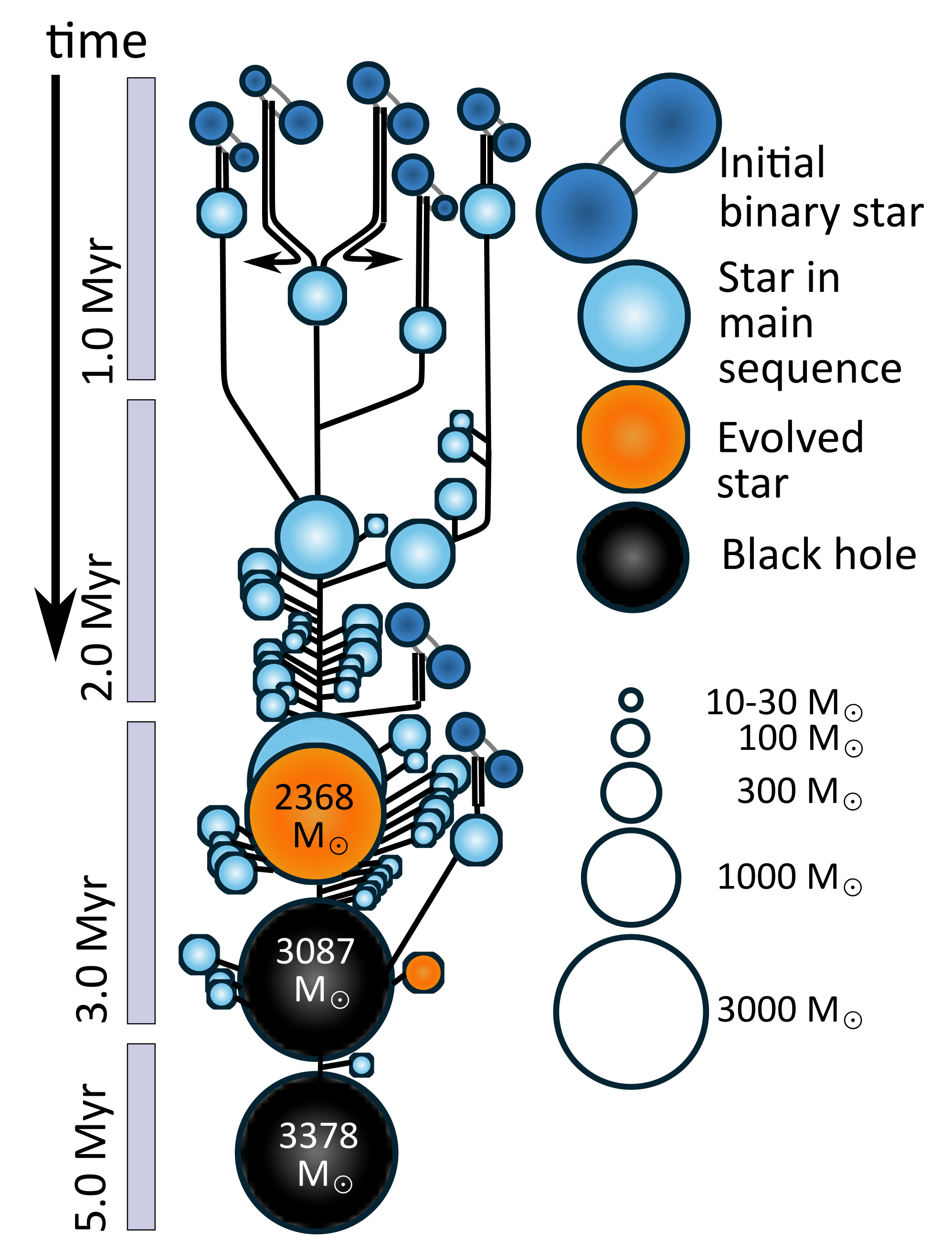}
\caption{The collisional formation of an IMBH with $M_\bullet=\msol{3378}$ in a single, dense isolated star cluster with $M_\mathrm{cl} = \msol{2.3\times10^5}$ and an initial binary population. Only collisions with secondary masses above $\msol{10}$ are displayed. Collisions of binary components and binary-binary interactions initiate the cascade of stellar collisions. The several collision products merge with the most massive collisionally growing star before forming BHs. At $t=5$ Myr, the cluster contains a single BH in the IMBH mass range.}
\label{fig: 1}
\end{centering}
\end{figure}

\section{Methods and initial conditions}

We use an updated GPU accelerated direct summation N-body code \bifrost{} \citep{Rantala2023} which uses the hierarchical variant \citep{Rantala2021} of the fourth order forward symplectic integrator \citep{Chin1997,Chin2005,Dehnen2017a}. Small scale few-body dynamics is treated using secular and regularised integrators (e.g. \citealt{Rantala2020}) with post-Newtonian (up to order PN3.5) equations of motion for BHs. We follow BH binary dynamics down to $\sim10$ Schwarzschild radii, and merging BHs receive relativistic GW recoil kicks based on the fitting formulas of \cite{Zlochower2015}. Timestep and particle neighbour assignment is sped up with an octree structure. Single and now also binary stellar evolution is followed in the simulations using the fast stellar population synthesis code \sevn{} \citep{Iorio2023}. We do not include stellar evolution prescriptions specific for triple stars. We assume no mass loss in stellar collisions, but also no explicit rejuvenation to avoid artificially prolonging the lifetimes of the stellar collision products. The simulation code and its recent updates are described in detail in \cite{Rantala2024b} and in a forthcoming study (FROST-CLUSTERS-II, Rantala et al., to be submitted). Throughout this study we use a metallicity of $Z=0.0002=0.01\zsol{}$.

Our initial setup for the isolated star cluster and hierarchical star cluster assembly simulations closely follows our models in \cite{Rantala2024b}. Besides single stars, we now also include populations of initial binary and triple stars in the clusters. The initial primary mass dependent binary and triple star fractions $f_\mathrm{bin}$ and $f_\mathrm{trip}$ as well as their orbital elements follow \cite{Offner2023}, \cite{Moe2017} and \cite{Winters2019}. For the binary population our setup closely resembles the implementation of \cite{Cournoyer-Cloutier2021}. Most of the massive stars ($f_\mathrm{bin}>0.9$ above $\msol{10}$) are initially in binaries and the initial global binary fraction is $f_\mathrm{bin}\sim0.29$. In the models with triples we initially have $f_\mathrm{trip}\sim0.05$. Single stars follow the \cite{Kroupa2001} initial mass function (IMF) down to $\msol{0.08}$ with a cluster mass dependent \citep{Weidner2006} initial mass function (IMF) cut-off $m_\mathrm{max,0}$ detailed in section 3.2 of \cite{Rantala2024b}.

For individual star clusters with $1.7 \times 10^4 \leq N \leq 10^6$ stars ($ \msol{9.4\times10^3}\leq M_\mathrm{cl}\leq\msol{6\times10^5}$) we assume the \cite{Plummer1911} density profile with small cluster birth radii (e.g. \citealt{Marks2012}, Eq. 11 of \citealt{Rantala2024b}) resulting in maximum initial stellar densities up to $\rhosol{3\times10^6}$ and surface densities $\sim$ a few times $\Sigmasol{10^5}$, close to the properties of the early $z\sim10$ JWST proto globular clusters \citep{Adamo2024}. We set up $10$ random realisations for each of $10$ different cluster masses all including initial binary stars. We run the $100$ isolated models for $t=5$ Myr as in \citealt{Rantala2024b}.

For the hierarchical models we consider setups with three different stellar multiplicity properties (singles only, including binaries, including binaries and triples), generating three random realisations for each setup. The nine hierarchical cluster assembly regions consist of up to $900$ star clusters sampled from the universal slope $-2$ power law star cluster mass function (e.g. \citealt{Elmegreen1996}) up to maximum sub-cluster mass of $M_\mathrm{cl}=\msol{2.5\times10^5}$. The most massive cluster is placed at the origin at zero velocity while the infalling clusters inhabit a spherical volume of $r=50$ pc in radius with $v_\mathrm{r} = -3.5$ km/s and a random velocity component of the same magnitude (see \citealt{Rantala2024b} for details). The total mass of each hierarchical model is $\sim \msol{1.0\times10^6}$ with $N\sim1.8\times10^6$ stars in the initial conditions. The hierarchical models are run until $t=10$ Myr after which most sub-clusters have merged into a single cluster typically containing multiple IMBHs in the mass range of $M_\bullet \sim \msol{10^3}$--$\msol{10^4}$.

\section{The crucial role of initial binary stars for collisional IMBH formation}

\subsection{Isolated cluster models}

We show a merger tree for a representative collisionally growing star in Fig. \ref{fig: 1} resulting in an IMBH of $M_\bullet=\msol{3378}$ in an isolated star cluster model with $M_\mathrm{cl}=\msol{2.3\times10^5}$ and an initial binary star population. In models without initial binaries, typically only one star per isolated cluster can efficiently grow through collisions (e.g. \citealt{Baumgardt2011,Fujii2013,Rantala2024b}). Initial binaries ($f_\mathrm{bin}\geq0.1$) can allow two or more stars per cluster to collisionally grow in Monte Carlo models \citep{Gurkan2006}. Pop-III star clusters may occasionally result in massive stellar binaries (e.g. \citealt{Reinoso2025}). \cite{Wang2022} and \cite{Wang_Hanzhang2025} found two IMBHs per Pop-III star cluster in their N-body models assuming $f_\mathrm{bin}=1$. In our isolated models we rarely find several IMBHs per cluster. Several massive stars may each experience multiple collisions, but they almost always merge with the most massive collision product before reaching the end of their lives. 

\begin{figure}
\includegraphics[width=0.9\columnwidth]{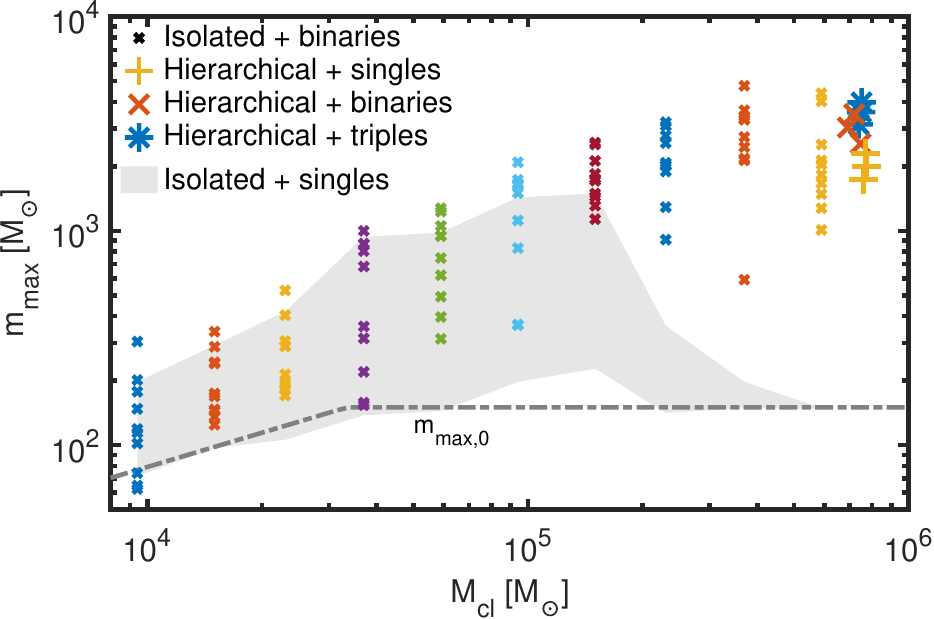}
\caption{The maximum mass $m_\mathrm{max}$ of stellar collision products in isolated clusters with initial binaries (small crosses) and hierarchical (large symbols) setups as a function of star cluster mass $M_\mathrm{cl}$. The shaded area shows the isolated single star models of \citet{Rantala2024b}. With initial stellar binaries there is no suppression of collisions at $M_\mathrm{cl}\gtrsim1.5\times \msol{10^5}$. Stellar multiplicity in the models results in higher masses of stellar collision products by a factor of up to $\sim2$--$3$.}
\label{fig: 2}
\end{figure}

Initial stellar binaries are crucial for driving collisional IMBH formation in dense, massive star clusters with high velocity dispersions (Fig. \ref{fig: 2}). Isolated single star cluster models with $M_\mathrm{cl} \lesssim \msol{1.5\times10^5}$ can collisionally grow stars up to $m_\mathrm{max}\sim0.02 M_\mathrm{cl}$. In higher mass clusters the collisional growth in single star models is strongly suppressed as high velocity dispersions inhibits \citep{Goodman1993} the dynamical binary formation \citep{Rantala2024b} and due to the competition of the cluster core collapse timescale and massive star lifetimes of $\lesssim3$ Myr (e.g. \citealt{PortegiesZwart1999}). When a population of initial binary stars is included in the simulations, the velocity dispersion bottleneck vanishes, and more massive isolated clusters form increasingly massive stars through repeated collisions. This highlights the role of binary systems in the first collisions of the runaway collision cascades \citep{Gaburov2008} and the increased collision cross sections in single-binary and binary-binary interactions \citep{Fregeau2004}. The most massive collisionally grown star in the isolated models with binaries reaches $M_\bullet \sim \msol{4000}$, a factor of $\sim2.7$ higher compared to the models without initial binaries.

\subsection{Hierarchically assembling cluster models}
The three new hierarchical single star models are very similar to the hierarchical runs of \cite{Rantala2024b}. $3$-$7$ IMBHs form per assembled cluster, and the most massive IMBHs in each random realisation reach masses of $M_\bullet= \msol{1809}$, $\msol{2708}$ and $\msol{2671}$. Several IMBHs are ejected from the clusters via strong Newtonian few-body interactions with the other seeds, and at $t=10$ Myr two of the single star models still contain two or more IMBHs.

\begin{figure*}
\includegraphics[width=0.7\textwidth]{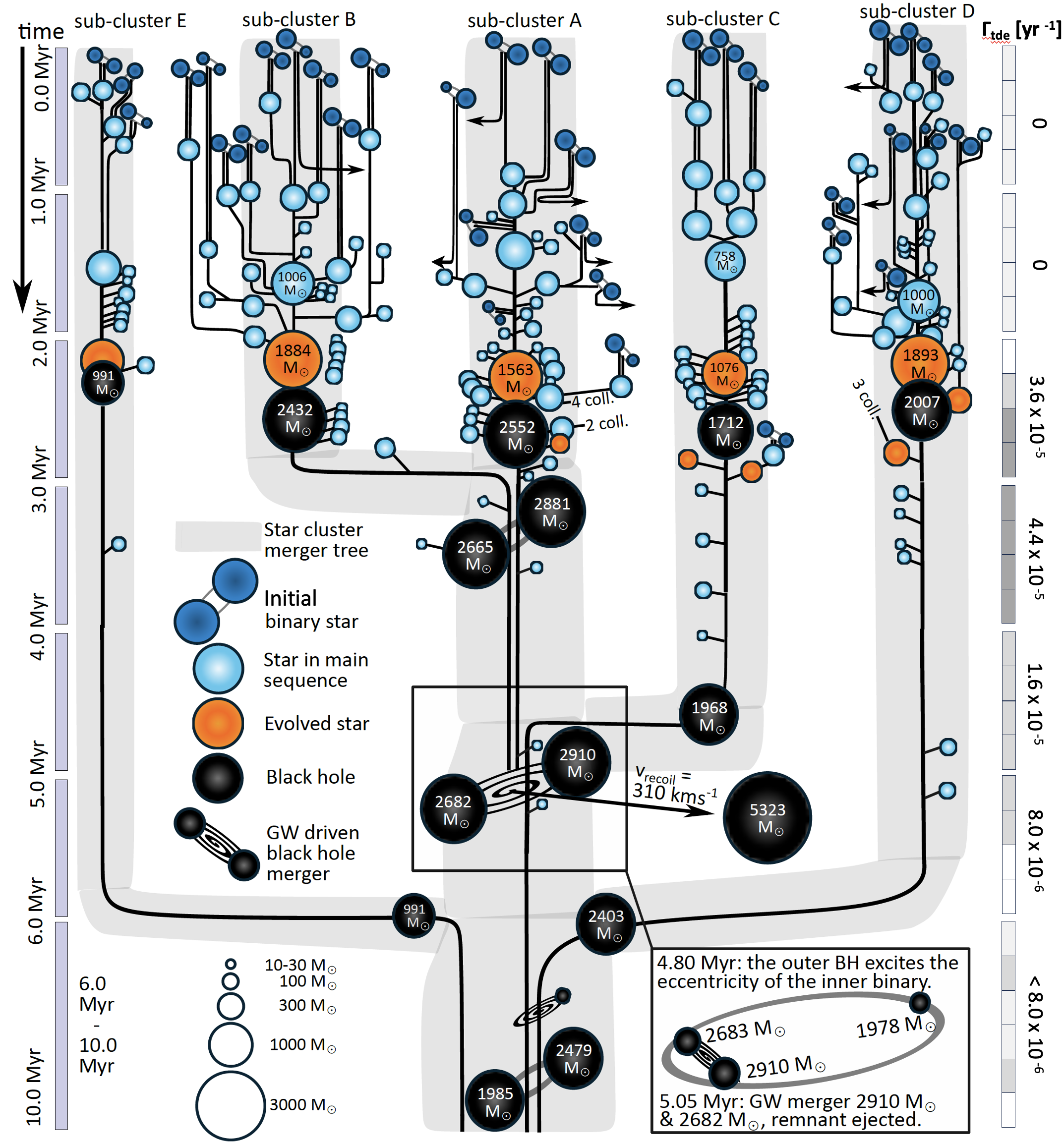}
\caption{An illustration of stellar, sub-cluster and IMBH merger trees in a hierarchical cluster assembly simulation with initial binaries. Only collisions with stars above $\msol{10}$ are shown. Runaway stellar collisions initiated by binary interactions rapidly ($t\lesssim2.5$ Myr) lead to formation of stars with masses up to $\sim\msol{2600}$ in the central growing cluster and four infalling sub-clusters. These stars end their lives forming IMBHs, and further grow by up to $10$--$15\%$ by tidally disrupting stars. IMBH mass binaries with close to equal mass ratios form twice in this simulation, and the first such binary with $M_\bullet=\msol{2910}$ and $M_\bullet=\msol{2682}$ is driven to a GW merger by a hierarchical IMBH triple interaction. The merger remnant ($M_\bullet\sim\msol{5300}$) is ejected from the cluster due to a GW recoil kick. Nevertheless, three IMBHs with masses up to $M_\bullet\sim \msol{2500}$ occupy the assembled cluster at the end of the simulation at $t=10$ Myr. The vertical bar on the right indicates the maximum TDE rate $\Gamma$ by IMBHs each Myr calculated in intervals of $0.25$ Myr (short bar sections).}
\label{fig: 3}
\end{figure*}

\begin{figure}
\includegraphics[width=\columnwidth]{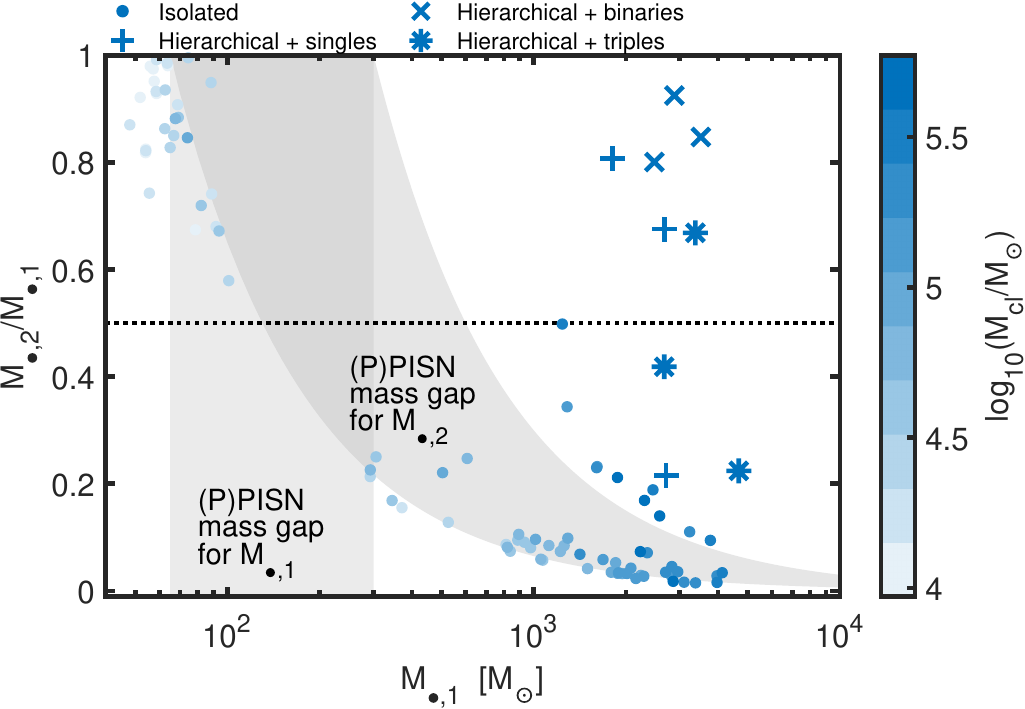}
\caption{The most massive BH masses $M_\mathrm{\bullet,1}$ and mass ratios ($q=M_\mathrm{\bullet,2}/M_\mathrm{\bullet,1}$) of two most massive IMBHs formed in the isolated (small filled circles) and hierarchical (large symbols) cluster simulations. Note that the two BHs do not necessarily form a binary in the runs. The second most massive BHs in the isolated runs are mostly below the pulsational pair-instability ((P)PISN) mass gap with a small number of massive secondary BHs reaching $0.2 \lesssim q \lesssim 0.5$ with $M_{\bullet,1} > \msol{1000}$. Meanwhile, 6 out of 9 hierarchical cluster assembly models have $q>0.65$ in this mass range.}
\label{fig: 4}
\end{figure}

\begin{figure*}
\includegraphics[width=0.7\textwidth]{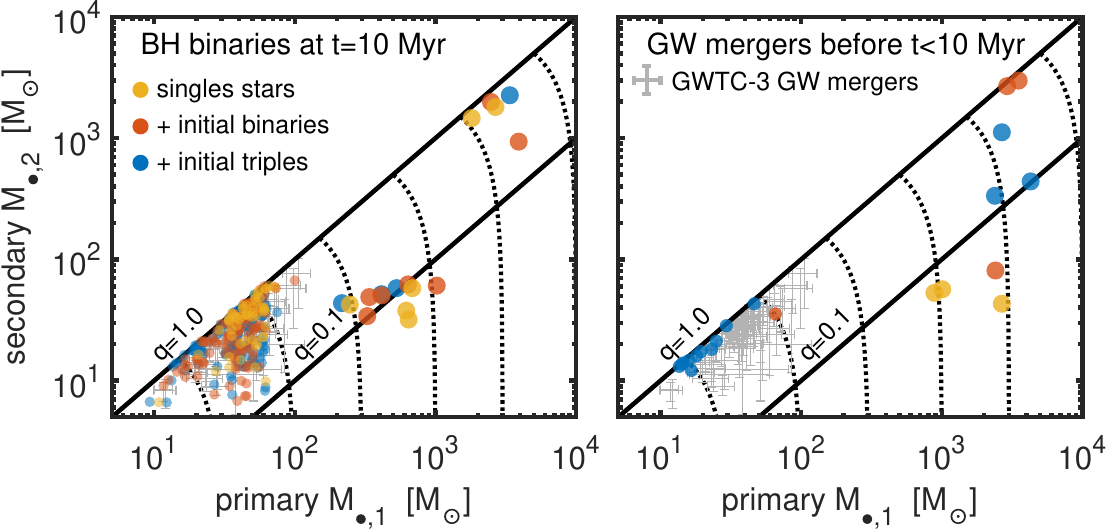}
\caption{The primary and secondary masses of BH binaries at $t=10$ Myr (left panel) and the GW mergers before $t<10$ Myr (right panel) in the hierarchical cluster assembly models. The Advanced LIGO / Advanced Virgo GW mergers in the GWTC-3 catalogue is shown as the cross symbols. Besides the GWTC-3 like stellar BH population, our hierarchical models result in IMBH-BH binaries and IMBH-BH GW mergers with $\msol{300} \lesssim M_\mathrm{\bullet,1} \lesssim \msol{3000}$ and $0.03 \lesssim q \lesssim 0.1$. Most importantly, the hierarchical cluster assembly drives the formation of close to equal mass IMBH binaries and GW mergers with $M_\mathrm{\bullet,1} \gtrsim \msol{2500}$ in less than $t=10$ Myr. The dashed lines indicate logarithmically spaced fixed values for the total binary mass $M_\mathrm{\bullet,1}+M_\mathrm{\bullet,2}$.}
\label{fig: 5}
\end{figure*}

For the more realistic hierarchical models including initial binaries, the merger trees of growing stars, IMBHs and their host clusters are illustrated in Fig. \ref{fig: 3} for one of the models. Five sub-clusters form IMBHs above $M_\bullet \gtrsim \msol{1000}$ through stellar collisions. In contrast to the single star setups, initial binary and triple populations allow for the formation in the most massive central star clusters as well. Three sub-clusters above the threshold mass of $M_\mathrm{cl}=\msol{10^4}$ above which collisions can frequently occur do not form BHs in the IMBH mass range. For all binary and triple models the first stellar collisions in the clusters occur early, before $t=0.5$ Myr, and almost always involve the members of a initial binary system, either colliding with each other or with other stars. By $t\sim2.5$ Myr, the five sub-clusters have formed IMBHs in the range of $\sim \msol{1000} \lesssim M_\bullet \lesssim \msol{2550}$. 

After their formation, the IMBHs may subsequently grow by tidally disrupting (mostly) main sequence stars as well as evolved stars. Tidal disruption events can contribute up to $10\%$--$15\%$ of the IMBH mass with most of the TDEs occurring soon after the formation of the IMBHs. The number of TDEs by IMBHs per assembled cluster ($\sim40$--$50$ for $M_\bullet\gtrsim \msol{100}$ and ($\sim30$--$45$ for $M_\bullet\gtrsim \msol{1000}$) in our models show no strong dependence on initial stellar multiplicity. The rate of TDEs by $M_\bullet \geq \msol{100}$ IMBHs peaks at $\Gamma\sim 4$--$5\times10^{-5}$ yr$^{-1}$ per assembling cluster at $2.5$ Myr $\lesssim t \lesssim 3.75$ Myr. At later times, the central density of the formed cluster decreases due to mass loss and dynamical relaxation and the TDE rate accordingly declines by a factor of $5$--$10$ (see also \citealt{Rizzuto2023}). During the $10$ Myr of the simulations, IMBH growth by GW mergers with stellar mass BHs is less important with only few ($<5$) IMBH-BH merger occurring in total in the models of this study.

IMBHs formed in the infalling sub-clusters rapidly sink to the main assembling cluster by the hierarchical star cluster merger process in $\lesssim10$ Myr, of the order of $\sim 1$ dynamical time $t_\mathrm{dyn}\sim5.3$ Myr of the initial system (see also \citealt{Shi2024}). In total 5 IMBHs formed through stellar collisions end up in the assembled cluster or its outskirts before $t=10$ Myr. A hierarchical triple IMBH (see also \citealt{Liu2024}) interaction ($M_\bullet=\msol{2910}, \msol{2682}$ and $\msol{1968}$) at $5$--$6$  Myr drives the most massive IMBH binary to a GW merger, the resulting BH with $M_\bullet \sim \msol{5300}$ being ejected from the cluster with a GW recoil kick of $v_\mathrm{kick}=310$ km/s. IMBH-IMBH mergers may provide an important channel for the early growth of SMBH seeds without any AGN or TDE signature, but only if the seeds can be retained in their environments. Around $t=6$ Myr, two star clusters deliver the subsequent IMBHs of $M_\bullet = \msol{2479}$ and $M_\bullet = \msol{991}$ into the central cluster. By the end of the simulation at $t=10$ Myr, the more massive of these has sunk to the centre of the main cluster, forming a binary system with the pre-existing central IMBH ($M_\bullet=\msol{1985}$). The future GW merger and GW recoil of these two IMBHs determines whether the cluster will eventually contain an IMBH of $M_\bullet \sim \msol{4000}$, $\sim \msol{1000}$, or no IMBHs at all.

Overall, the models with initial binaries form $4$--$6$ IMBHs per assembled cluster with on average $2$--$3$ times higher masses compared to the single star models. In the second hierarchical model with initial binaries several IMBHs form and one IMBH with $M_\bullet=\msol{542}$ is ejected from the final cluster in a strong Newtonian few-body interaction. The central IMBH binary at $t=10$ Myr has a mass ratio of $q \sim 1/4$ with IMBH masses of $M_\bullet=\msol{935}$ and $M_\bullet=\msol{3904}$. Another IMBH with $M_\bullet=\msol{638}$ from a cluster merger is slowly sinking at the outskirts of the cluster. The third hierarchical model with binaries produces the most massive GW merger of this study with close to equal IMBH masses of $M_\bullet=\msol{3516}$ and $M_\bullet=\msol{2980}$. The remnant IMBH with $M_\bullet=\msol{6190}$ became marginally unbound of the host cluster by the relativistic GW recoil kick of $v_\mathrm{kick}=107$ km/s. At $t=10$ Myr, three less massive IMBHs ($\msol{314} \lesssim M_\bullet \lesssim \msol{545}$) sink in the cluster outskirts, originating from three individual cluster mergers, but have not yet had a sufficient time to reach the cluster centre. 

Finally, we focus on hierarchical clusters with initial triples. The total number of any stellar collisions is the highest in the triples models, and the triple models give rise to $9$-$11$ IMBHs per assembled cluster. Still, including initial triples in the models only moderately increases the final collision product masses, the most massive IMBH in each triple simulations reaching $M_\bullet=\msol{3374}$, $\msol{4676}$ and $\msol{2479}$. $\sim30\%$ of the IMBHs are ejected before $t=10$ Myr via Newtonian interactions and GW recoil kicks. Initial stellar multiplicity results in earlier first collisions and a head start for the collision cascades compared to the single star models, leading to by a factor of $2$--$3$ higher IMBH masses compared to models with no initial multiplicity. Further increased stellar multiplicity  (triples vs binaries) results in a higher number of formed IMBHs per final cluster, but only mildly increases the IMBH masses.

\section{A characteristic GW fingerprint of collisional SMBH seed formation}

GW mergers of IMBHs ($M_\bullet \gtrsim \msol{1000}$) with similar masses ($q=M_\mathrm{\bullet,2}/M_\mathrm{\bullet,1}>0.5$) occur in the nine hierarchical models, but not in the 100 isolated cluster simulations. We show the masses of the first and second most massive (IM)BH of each simulation in Fig. \ref{fig: 4}. Note that these do not necessarily form a binary. The majority of the isolated models with $M_\mathrm{\bullet,1}\gtrsim \msol{1000}$ have $q\lesssim0.1$ with $M_\mathrm{\bullet,2}$ below or close to the lower edge of the pulsational pair-instability ((P)PISN) mass gap. Nearly $10$ isolated models have their second most massive BHs above the (P)PISN mass gap with $q\lesssim0.23$ for all but two models. Only two models of the $44$ isolated runs with $M_\mathrm{\bullet,1}\gtrsim \msol{1000}$ have more intermediate mass ratios, $q=0.34$ and $q=0.50$. On the other hand, $6/9$ of the hierarchical runs have $q\gtrsim0.65$ as multiple sub-clusters provide IMBHs to the final assembled cluster through the hierarchical merging process.

We present the mass ratios of all black hole binaries at $t = 10$ Myr and the already merged black holes the hierarchical simulations in the left and right panels of Fig. \ref{fig: 5}, respectively. In the stellar mass BH range all low mass BH binaries coincide with the primary and secondary BH masses from the GWTC-3 catalogue  by the Advanced LIGO and Advanced Virgo \citep{Abbott2023}. Due to the relatively short simulation time (see e.g. \citealt{Rizzuto2022} for comparison), there are only 12 stellar BH mergers before $t=10$ Myr, most of them in the triple setups in which the outer bodies of triple systems can drive the inner binaries to earlier mergers.

At $t=10$ Myr, the hierarchical simulations also host a population of BH binaries with $q\sim0.1$ consisting of one IMBH in the range of $\msol{300} \lesssim M_\mathrm{\bullet,1} \lesssim \msol{1000}$ and a stellar BH. At the higher IMBH masses up to $\sim \msol{3000}$, some IMBH-BH binaries have already merged within $10$ Myr (right panel of Fig. \ref{fig: 5}). Most interestingly, the hierarchical models form binaries of similar mass with $M_\mathrm{\bullet,1} \gtrsim \msol{2400}$ at $t=10$ Myr and a few IMBH-IMBH binaries with $M_\mathrm{\bullet,1} \gtrsim \msol{2400}$ and $q>0.1$ have already merged, two of them with mass ratios near unity. Therefore, in the hierarchical star cluster assembly scenario, multiple IMBHs can end up in the same assembling star cluster and they can also rapidly merge with each other within $10$ Myr of cluster evolution. The binary IMBH mergers are often driven by a third IMBH. This additional feature of the hierarchical assembly scenario leads to more rapid ($<10$ Myr) IMBH mergers than in timescale considerations of two IMBHs per cluster \citep{Amaro-Seoane2006,Amaro-Seoane2010,ArcaSedda2019,Rasskazov2020,Souvaitzis2025} driven by stellar hardening alone. In addition, triple IMBH interactions in the hierarchical cluster assembly models may also more often result in above near-zero eccentricities at GW mergers.

Third-generation ground-based GW detectors like the Einstein Telescope and the Cosmic Explorer will be sensitive to IMBH-IMBH mergers with masses $M_\mathrm{bin} = M_\mathrm{\bullet,1}+M_\mathrm{\bullet,2} \lesssim \msol{10^3}$ at $z<10$ \citep{Reali2024,EinsteinTelescopeCollaboration2025}. The $M_\mathrm{bin} \lesssim \msol{10^3}$ $z<10$ GW mergers most probably occur too late to belong the parent population of SMBHs powering the $z=6$--$7$ active galactic nuclei \citep{Fan2023,Inayoshi2020}, but can nevertheless constrain the mass function of the seeds that fail to grow at high redshifts. The sensitivity of LISA for equal-mass BH binaries at high redshifts rapidly increases from $M_\mathrm{bin}= \msol{10^3}$ towards $M_\mathrm{bin}=\msol{10^4}$ \citep{Amaro-Seoane2012}. Still, only our most massive IMBH binaries could be detectable with LISA at $z>10$. Overall, we expect that IMBH binaries with $M_\mathrm{bin} \gtrsim \msol{10^4}$ formed during hierarchical assembly of clusters with $M_\mathrm{cl} \gtrsim$ a few times $\msol{10^6}$ will be promising GW targets for LISA high redshifts. Lower-mass ($\lesssim\msol{10^3}$) $z>10$ IMBH mergers will be observed using GW space observatories with shorter interferometer arms such as TianQin \citep{Luo2016}.

\section{Summary}

We show that hierarchical massive star cluster formation at low metallicities ($\zsol{0.01}$) can lead to the formation of IMBHs and the rapid GW driven IMBH-IMBH mergers. The N-body models include evolving single stars and initial binary and triple systems as well as post-Newtonian dynamics for BHs. Initial stellar multiplicity leads to earlier stellar mergers, higher collision rates and a factor of $2$--$3$ larger IMBH masses compared to single star models. The hierarchical cluster assembly models frequently result in the formation of close to equal mass IMBH binaries and GW driven IMBH mergers. Each sub-cluster with $M_\mathrm{cl} \gtrsim \msol{10^4}$--$\msol{10^5}$ may provide a single IMBH for the hierarchically assembled cluster, often resulting in more than two IMBHs in the assembled cluster. In general, models with increasing stellar multiplicity lead to more SMBH seeds in the final cluster. In idealised isolated clusters, even initial stellar multiplicity does not result in IMBH-IMBH mergers.

The hierarchical scenario provides a rapid and efficient channel for IMBH formation through stellar collisions and tidal disruption events, their pairing in massive binaries via star cluster mergers, and finally to potential GW mergers driven by stellar hardening and triple IMBH interactions. The entire process can occurs in less than $10$ Myr. If retained in their formation environments, the IMBHs formed through repeated stellar collisions constitute a natural candidate for the seeds of SMBHs. SMBH seeds up to $\sim \msol{6200}$ ejected by GW recoil kicks often have relatively low velocities, only up to a few times $100$ km/s. While these kick velocities are high enough to unbind the IMBHs from their host clusters but not from their host galaxies. IMBH-IMBH mergers may constitute an important growth channel for the early SMBH seeds without any AGN or TDE signature. Our results suggest that a $M_\bullet \gtrsim \msol{1000}$ GW merger with similar IMBH masses is a characteristic GW fingerprint of star cluster assembly with simultaneous SMBH seed formation at high redshifts. At lower redshifts the IMBH formation and GW mergers may still proceed in dense, low metallicity environments. Besides GWs, tidal disruption events with TDE rates up to $\Gamma \sim 4$--$5\times10^{-5}$ yr$^{-1}$ per assembling star cluster might reveal a population of such low redshift IMBHs in the upcoming extragalactic wide field time domain surveys.

Other astrophysical channels may also lead to IMBH-IMBH mass GW mergers. Assuming only one IMBH forms per star cluster, these IMBH can still merge in later star cluster mergers \citep{Souvaitzis2025} e.g. in the galactic disks \citep{ArcaSedda2019} or in galactic centres (e.g. \citealt{Elmegreen2008, Dekel2025}). However, the sinking process itself or the subsequent mergers may be slow, lasting $\gg10$ Myr (e.g. \citealt{Wirth2020,Partmann2024}) especially if the SMBH seed is not embedded in its birth star cluster \citep{Zhou2025,Mukherjee2025}. Pristine metal-free Pop-III star clusters may also be a source of GWs from IMBH mergers (\citealt{Wang2022} and \citealt{Wang_Hanzhang2025}). The central IMBHs of dwarf galaxies may merge in the aftermath of dwarf galaxy mergers \citep{Tamfal2018,Bellovary2019,Khan2021} although the merger timescales are likely far longer than $10$ Myr unless the formed IMBH binaries have an eccentricity of near unity. Finally, IMBHs may grow, pair and merge in the disks of active galactic nuclei \citep{McKernan2012,McKernan2014}. However, this IMBH merger channel relies on the pre-existing SMBH and its accretion disk, so it can operate only after sufficiently massive SMBHs have formed.

In this work we focused on low metallicity models with $Z=\zsol{0.01}$ which results in relatively weak stellar wind mass loss from the collisionally grown stars. However, it is not uncommon for JWST $z\sim10$ galaxies and AGN (e.g. \citealt{Maiolino2024}) such as GN-z11 to show $Z \sim \zsol{0.1}$ \citep{Bunker2023} or even higher metallicities (e.g. \citealt{Isobe2023}). In the future we plan to extend our models into higher metallicities and examine whether dense star formation can produce IMBHs and IMBH-IMBH GW mergers in the early enriched $z\sim10$ environments, in clumpy galaxies at lower redshifts (e.g. \citealt{Genzel2011}), and in the local Universe.

\section*{Data availability statement}
The data relevant to this article will be shared on reasonable request to the corresponding author.

\section*{Acknowledgments}
The numerical simulations were performed using facilities in Germany hosted by the Max Planck Computing and Data Facility (MPCDF) and the JUWELS Booster of the Jülich supercomputing centre (GCS project 59949 frost-smbh-origins). TN acknowledges support from the Deutsche Forschungsgemeinschaft (DFG, German Research Foundation) under Germany's Excellence Strategy - EXC-2094 - 390783311 from the DFG Cluster of Excellence "ORIGINS". The development of the SEVN code was enabled by M. Mapelli’s ERC Consolidator grant DEMOBLACK by the European Research Council under contract no. 770017.


\bibliographystyle{mnras}
\interlinepenalty=10000
\bibliography{manuscript}




\bsp	
\label{lastpage}
\end{document}